\documentclass[showpacs,amsmath,amssymb,pre]{revtex4}
\def\baselinestretch{2.0}
\usepackage{bm}
\usepackage{epsfig}
\usepackage{color}
\usepackage{natbib}
\usepackage{graphicx}

\begin{document}

\begin{large}

\title{Standard and inverse site percolation of straight rigid rods on triangular lattices: Isotropic and nematic deposition/removal}

\author{L. S. Ramirez$^{1,\dag}$, P. M. Pasinetti$^1$, W. Lebrecht$^{2}$, A.J. Ramirez-Pastor$^1$}
\affiliation{$^1$Departamento de F\'{\i}sica, Instituto de F\'{\i}sica Aplicada, Universidad Nacional de San Luis-CONICET, Ej\'ercito de Los Andes 950, D5700HHW, San  Luis, Argentina}
\affiliation{$^2$Departamento de F\'{\i}sica, Universidad de La Frontera, Casilla 54-D, Temuco, Chile}

\begin{abstract}
Numerical simulations and finite-size scaling analysis have been carried out to study standard and inverse percolation of straight rigid rods on triangular lattices. In the case of standard percolation, the lattice is initially empty. Then, linear $k$-mers (sets of $k$ linear consecutive sites) are randomly and sequentially deposited on the lattice. In the case of inverse percolation, the process starts with an initial configuration where all lattice sites are occupied and, consequently, the opposite sides of the lattice are connected by nearest-neighbor occupied sites. Then, the system is diluted by randomly removing linear $k$-mers from the lattice. Two schemes are used for the depositing/removing process:  isotropic scheme, where the deposition (removal) of the linear objects occurs with the same probability in any lattice direction; and nematic scheme, where one lattice direction is privileged for depositing (removing) the particles. The study is conducted by following the behavior of four critical concentrations with the size $k$: $(i)$[$(ii)$] standard isotropic[nematic] percolation threshold $\theta_{c,k}$[$\vartheta_{c,k}$], which represents the minimum concentration of occupied sites at which an infinite cluster of occupied nearest-neighbor sites extends from one side of the system to the other. $\theta_{c,k}$[$\vartheta_{c,k}$] is reached by isotropic[nematic] deposition of straight rigid $k$-mers on an initially empty lattice; and $(iii)$[$(iv)$] inverse isotropic[nematic] percolation threshold $\theta^i_{c,k}$[$\vartheta^i_{c,k}$], which corresponds to the maximum concentration of occupied sites for which connectivity disappears. $\theta^i_{c,k}$[$\vartheta^i_{c,k}$] is reached after isotropically[nematically] removing straight rigid $k$-mers from an initially fully occupied lattice. $\theta_{c,k}$, $\vartheta_{c,k}$, $\theta^i_{c,k}$ and $\vartheta^i_{c,k}$ are determined for a wide range of $k$ ($2 \leq k \leq 512$). The obtained results indicate that: $(1)$ $\theta_{c,k}$[$\theta^i_{c,k}$] exhibits a non-monotonous dependence on the size $k$. It decreases[increases] for small particles sizes, goes through a minimum[maximum] around $k = 11$, and finally increases and asymptotically converges towards a definite value for large segments $\theta_{c,k \rightarrow \infty}=0.500(2)$[$\theta^i_{c,k \rightarrow \infty}=0.500(1)$]; 
$(2)$ $\vartheta_{c,k}$[$\vartheta^i_{c,k}$] depicts a monotonous behavior in terms of $k$. It rapidly increases[decreases] for small particles sizes and asymptotically converges towards a definite value for infinitely long $k$-mers $\vartheta_{c,k \rightarrow \infty}=0.5334(6)$[$\vartheta^i_{c,k \rightarrow \infty}=0.4666(6)$]; $(3)$ for both isotropic and nematic models, the curves of standard and inverse percolation thresholds are symmetric to each other with respect to the line $\theta = 0.5$. Thus, a complementary property is found $\theta_{c,k} + \theta^i_{c,k} = 1$ (and $\vartheta_{c,k} + \vartheta^i_{c,k} = 1$), which has not been observed in other regular lattices. This condition is analytically validated by using exact enumeration of configurations for small systems; and $(4)$ in all cases, the critical concentration curves divide the $\theta$-space in a percolating region and a non-percolating region. These phases extend to infinity in the space of the parameter $k$ so that the model presents percolation transition for the whole range of $k$.

\end{abstract}


\maketitle

\noindent $\dag$ To whom all correspondence should be addressed. E-mail: lsramirez@unsl.edu.ar

\newpage

\section{Introduction}
\label{introduccion}

Since its introduction in the 1950s by Hammersley and Broadbent \cite{Hammersley,Broadbent}, the percolation problem has been a focal point of statistical mechanics and it has been applied to a wide range of phenomena in physics, chemistry, biology, and materials science where connectivity and clustering play an important role \cite{Stauffer,Sahimi,Dorogo,Newman,Newman1,Cohen,Kornbluth,Lowinger,Gao,Coniglio,Kenah,Yazdi,Chatter,Tara2018a}. Percolation theory has also provided insight into the behavior of more complicated models exhibiting phase transitions and critical phenomena \cite{Stauffer,Sahimi,Grimmett,Christensen,Bollobas}.

Usually, the percolation model in a lattice is classified into two categories, namely, site model and bond model \cite{Stauffer}. In the site [bond] model, sites [bonds] of a lattice are randomly occupied with a probability $\theta$
or empty (nonoccupied) with a probability $1-\theta$. Nearest-neighboring occupied sites (bonds) form structures called clusters. In the limit of an infinite lattice, there is a well-defined value of $\theta$, known as percolation threshold $\theta_c$, at which an infinite cluster extends from one side of the system to the other. The percolation transition is then a geometrical phase transition where the critical concentration separates a phase of finite clusters from a phase where a macroscopic, spanning, or infinite cluster is present. The exact determination of $\theta_c$ is an unsolved problem except for a few cases.

An interesting phenomenon occurs when the lattice is occupied by extended objects (objects occupying more than one lattice site). Under these conditions, the final state generated by irreversible adsorption is a disordered state (known as jamming state), in which no more objects can be deposited due to the absence of free space of appropriate size and shape \cite{Feder,Evans}. The corresponding limiting or jamming coverage, $\theta_j$, is less than that corresponding to close packing ($\theta_j <1$). Thus, the jamming coverage has an important role in the determination of the percolation threshold, and the interplay between jamming and percolation is relevant for the description of various deposition processes.

One of the simplest processes that produces a jamming state is the random sequential adsorption (RSA) of straight rigid $k$-mers (objects occupying $k$ consecutive sites along one of the lattice directions) on infinite two-dimensional (2D) lattices. In RSA processes, particles are randomly, sequentially and irreversibly deposited onto a substrate without overlapping each other \cite{Feder,Evans,Cadilhe1,Cadilhe2}. 

In the case of straight rigid $k$-mers on triangular lattices, which is the focus of this article, Budinski-Petkovi\'c and Kozmidis-Luburi\'c \cite{Budi1997} examined the kinetics of the RSA for values of $k$ between 1 and 11 and lattice
size $L = 128$. The coverage of the surface and the jamming limits were calculated by Monte Carlo simulations. The authors found that the jamming coverage decreases monotonically as the $k$-mer size increases. Later, Budinski-Petkovi\'c et al. \cite{Budi2012} investigated percolation and jamming thresholds for RSA of extended objects on triangular lattices. Numerical simulations were performed for lattices with linear size up to $L = 1000$, and objects of different sizes and
shapes (linear segments; angled objects; triangles and hexagons). It was found that for elongated shapes the percolation threshold monotonically decreases, while for more compact shapes it monotonically increases with the object size. In the
case of linear segments with values of $k$ up to 20, the obtained results revealed that (1) the jamming coverage monotonically decreases with $k$, and tends to 0.56(1) as the length of the rods increases; (2) the percolation threshold decreases for shorter $k$-mers, reaches a value $\theta_c \approx 0.40$  for $k = 12$, and, it seems that $\theta_c$ does not significantly depend on $k$ for larger $k$-mers; and (3) consequently, the ratio $\theta_c/\theta_j$ increases with $k$. The effects of anisotropy \cite{Budi2011} and the presence of defects \cite{Budi2016} on the jamming behavior were also studied by the group of Budinski-Petkovi\'c et al.

In the line of Refs. \cite{Budi1997,Budi2012,Budi2011,Budi2016}, three previous articles from our group \cite{JSTAT6,JSTAT8,PRE20} were devoted to the study jamming and percolation of straight rigid $k$-mers on triangular lattices. These papers will be referred to as Papers I, II and III, respectively. In Paper I, the results in Refs. \cite{Budi1997,Budi2012} were extended to larger lattices and longer objects: $L/k = 100, 150, 200, 300$
and $2 \leq k  \leq  128$ for jamming calculations, and $L/k = 32, 40, 50, 75, 100$ and $2  \leq  k  \leq  256$
for percolation analysis. The obtained results showed that the jamming coverage decreases monotonically approaching the asymptotic value of 0.5976(5) for large values of $k$.  On the other hand, a nonmonotonic $k$ size dependence was found
for the percolation threshold, in accordance with previous data for square lattices \cite{Leroyer,Kondrat,Tara2012,Slutskii}. In addition, a complete analysis of critical exponents revealed that the percolation phase transition involved in the system has the same universality class of the ordinary random percolation, regardless of the value of $k$ considered.

In Paper II, the problem of inverse percolation by removing straight rigid $k$-mers from 2D triangular lattices was investigated by using numerical simulations and finite-size scaling analysis. The study of inverse percolation problem starts with an initial configuration, where all lattice sites are occupied and, consequently, the opposite sides of the lattice are connected by nearest-neighbor occupied sites. Then, the system is diluted by randomly removing straight rigid rods $k$-mers from the surface. The main objective is to obtain the maximum concentration of occupied sites (minimum concentration of empty sites) at which the connectivity disappears. This particular value of the concentration is named
the inverse percolation threshold $\theta^i_c$, and determines a well-defined geometrical (second order) phase transition in the system. 

The results in Paper II, obtained for $k$ ranging from 2 to 256, revealed that (i) the inverse percolation threshold exhibits a nonmonotonic behavior as a function of the $k$-mer size: it grows from $k = 1$ to $k = 10$, goes through a maximum at $k = 11$, and finally decreases again and asymptotically converges towards a definite value for large values of $k$; (ii) the percolating and non-percolating phases extend to infinity in the space of the parameter $k$ and, consequently, the model presents percolation transition in all the ranges of $k$; and (iii) the phase transition occurring in the system belongs to the standard random percolation universality class regardless of the value of $k$ considered.

More recently, in Paper III, numerical simulations were used to study the percolation behavior of aligned rigid rods of length $k$ on 2D triangular lattices. The linear $k$-mers were irreversibly deposited along one of the directions of the lattice. The results, obtained for $k$ ranging from 2 to 80, showed that the percolation threshold exhibits a increasing function when it is plotted as a function of the $k$-mer size. This behavior is completely different from that observed for square lattices, where the percolation threshold decreases with $k$ \cite{PRE7}. In addition, an exhaustive study of critical exponents and universality was carried out, showing that the phase transition occurring in the system belongs to the standard random percolation universality class.

In this work, the problem of standard and inverse percolation of straight rigid $k$-mers isotropically deposited on 2D triangular lattices is revisited. The most important simulation results obtained in previous papers are briefly reviewed. In addition, the calculations are extended to longer $k$-mers (up to $k=512$) and values of $k$ in the range $[10,14]$, where the percolation threshold curves (standard and inverse) show a change in slope. The new findings allow us (1) to precisely determine the position of the minimum (maximum) observed in la curve of $\theta_c$ ($\theta^i_c$) as an function of $k$; and (2) to conclude that $\theta_c+\theta^i_c=1$ for all value of $k$. The simple relationship between $\theta_c$ and $\theta^i_c$ will also be validated by exact results for small lattices obtained by exact enumeration of states in configurational space. The result in point (2) would indicate that, in both standard and inverse, the percolating and non-percolating phases extend to infinity in the space of the parameter $k$.

The problem of aligned straight rigid $k$-mers deposited on triangular lattices is also studied. In the case of inverse percolation, the results of percolation threshold versus $k$ are presented for the first time in the literature. As in the isotropic case, the sum of standard and inverse percolation thresholds equals one, confirming the generality of this behavior in triangular lattices. On the other hand, the observed complementarity between standard and inverse percolation thresholds contributes to justify the striking increasing trend of the percolation threshold with $k$ found in Paper III. 

The rest of the paper is organized as follows. In Sec. \ref{isotro}, standard and inverse percolation of straight rigid $k$-mers on 2D triangular lattices is revisited. Calculations are extended to longer objects. In addition, numerical results are supplemented by exact results for small lattices coming from a complete enumeration of configurations. The problem of percolation of aligned rigid rods is addressed in Sec. \ref{nematic}. Finally, the conclusions are drawn in
Sec. \ref{conclusiones}.

\section{Percolation of straight rigid rods isotropically deposited on triangular lattices}\label{isotro}

In this section, we will revisit the percolation problem of straight rigid rods isotropically deposited on triangular lattices, this time focusing on the complementarity property of the standard and inverse percolation thresholds: $\theta_c+\theta^i_c=1$. For this purpose, new numerical simulations are presented in Sec. \ref{modiso} and an analytical approach is introduced in Sec. \ref{cluster}.

\subsection{Simulation results: dependence of the standard and inverse percolation thresholds on the $k$-mer size}\label{modiso}

The percolation problem is defined on a 2D triangular lattice. In the computer simulations, a rhombus-shaped system of $M = L \times L$ sites ($L$ rows and $L$ columns) is used [see Fig. \ref{fig1}]. Each site can be empty (hole) or occupied. Occupied and empty sites are distributed with a concentration $\theta$ and $\theta^*(= 1 - \theta)$, respectively. Nearest-neighbor occupied sites form structures called clusters, and the distribution of these sites determines the probability of the existence of a large cluster (also called ``infinite" cluster, inspired by the thermodynamic limit) that connects from one side of the lattice to the other. 

\begin{figure}
\includegraphics[width=0.95\columnwidth]{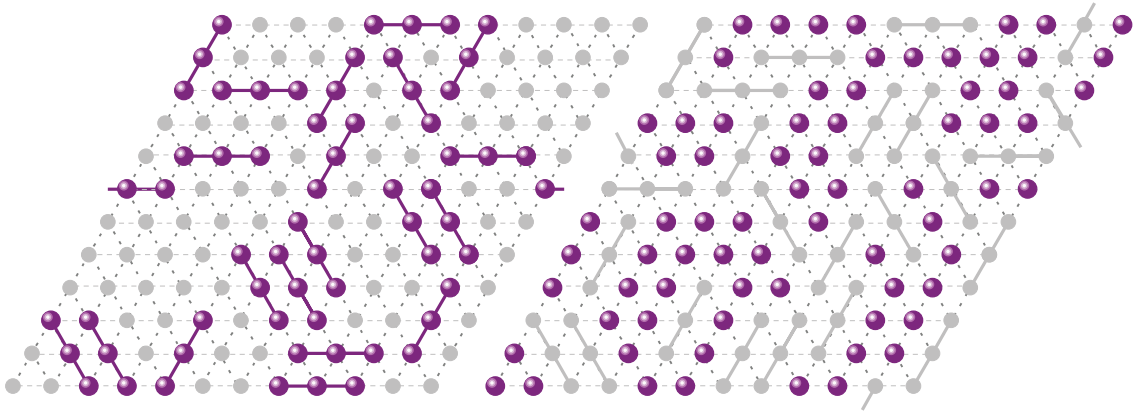}
\caption{(a) Schematic representation of a typical configuration obtained by isotropically depositing $3$-mers ($k=3$) on a $L \times L$ lattice with $L=12$. Solid spheres joined by lines represent the deposited $k$-mers and gray circles correspond to empty sites. (b) Schematic representation of a typical configuration obtained by removing $3$-mers from an initially fully occupied $L \times L$ lattice with $L=12$. Solid spheres represent occupied sites and grey circles indicate the empty sites resulting from the removal of the $k$-mers. Periodic boundary conditions are considered in parts (a) and (b).}  \label{fig1}
\end{figure}

Two procedures have been considered. In the first, straight rigid $k$-mers (with $k \geq 2$) are deposited randomly, sequentially and irreversibly on an initially empty lattice. This scheme, known as random sequential adsorption (RSA) \cite{Evans}, is as follows: (i) one of the three ($x_1, x_2, x_3$) possible lattice directions and a starting site are randomly chosen; (ii) if, beginning at the chosen site, there are $k$ consecutive empty sites along the direction selected in (i), then a $k$-mer is deposited on those sites (the $k$ sites are marked as occupied). Otherwise, the attempt is rejected. When $N$ rods are deposited, the concentration of occupied and empty sites is $\theta = kN/M$ and $\theta^* = (M - kN)/M$, respectively. 

In the second procedure, the process starts with a fully occupied lattice ($\theta$ = 1 and $\theta^*$ = 0). Then,
the system is diluted by randomly removing particles from the lattice. The mechanism of dilution is as follows: (i) a linear $k$-uple of $k$ consecutive sites is chosen at random; (ii) if the $k$ sites selected in step (i) are occupied by $k$ particles, then a $k$-mer is removed from those sites. Otherwise, the attempt is rejected. When $N$ rods are removed, the
concentration of particles (holes) is $\theta = (M - kN)/M$ ($\theta^* = kN/M$). 

In both first process (deposition) and second process (removal), periodic boundary conditions are considered. By using the first procedure (standard RSA), the lattice coverage is increased until finding a concentration at which a cluster of nearest-neighbor sites extends from one side to the opposite one of the system. This constitutes the so-called standard percolation problem, and the critical concentration rate is named standard percolation threshold. 

On the other hand, when the $k$-mers are removed from an initially fully occupied lattice (second procedure), the fraction of occupied sites decreases until reaching a concentration at which the connectivity disappears. The model of such a process can be thought of as an inverse percolation problem. The corresponding critical concentration is then named inverse percolation threshold. The term inverse is used simply to indicate that the size of the conductive phase diminishes during the removing process and the percolation transition occurs between a percolating and a nonpercolating state.

Typical configurations obtained from first and second procedures are shown in Figs.\ref{fig1}(a) and \ref{fig1}(b). A system with $L=12$ and $k=3$ is depicted in the figure. In part (a), solid spheres joined by lines represent the deposited $k$-mers and gray circles correspond to empty sites. In part (b), solid spheres represent occupied sites and gray circles indicate the empty sites resulting from the removal of the $k$-mers.  

In Papers I and II, standard and inverse percolation thresholds were calculated by using an extrapolation method based on scaling laws \cite{Stauffer}:
\begin{equation}
\theta_{c,k}(L)= \theta_{c,k} + A_k L^{-1/\nu},
\label{extrapolationdir}
\end{equation}
and
\begin{equation}
\theta^i_{c,k}(L)= \theta^i_{c,k} + A^i_k L^{-1/\nu},
\label{extrapolationinv}
\end{equation}
where the supra-index $i$ refers to the inverse percolation problem. Thus, $\theta_{c,k}[\theta^i_{c,k}]$ is the standard[inverse] percolation threshold in the thermodynamic limit ($L \rightarrow \infty$) for an object of size $k$; $A_k$ and $A^i_k$ are non-universal constants and $\nu$ is the critical exponent of the correlation length, which in two dimensions is $\nu = 4/3$ \cite{Stauffer}. The quantities $\theta_{c,k}(L)[\theta^i_{c,k}(L)]$ represent the percolation thresholds for finite lattices.

An standard method to obtain $\theta_{c,k}(L)[\theta^i_{c,k}(L)]$ consists of the following steps: (a) the construction
of a triangular lattice of linear size $L$ and coverage $\theta$, and (b) the cluster analysis using the Hoshen and Kopelman algorithm \cite{Hoshen}. A total of $r$ independent runs of such two steps procedure are carried out for each
lattice size $L$ and size $k$. From these runs, a number $r^*$ of them present a percolating cluster. Then, a percolation probability can be defined as $R_{L,k}(\theta)=r^*/r$. In the present study, open boundary conditions are used to determine the percolation quantities. 

In the case of standard[inverse] percolation problem, $R_{L,k}(\theta)$ is an increasing[decreasing] sigmoid function of the coverage, and $\theta_{c,k}(L)[\theta^i_{c,k}(L)]$ can be obtained from the position of the inflection point of the function $R_{L,k}(\theta)$. Interested readers are referred to Papers I-III for a more complete description of the technique to determine the percolation threshold from the percolation probability functions.

By following the procedure in Eqs. (\ref{extrapolationdir}) and (\ref{extrapolationinv}), standard and inverse percolation thresholds were calculated in Papers I and II, respectively. In Paper I, the simulations were carried out for values of $k$ ranging between 2 and 256 and lattice sizes $L/k = 32, 40, 50, 75, 100$. The corresponding values used in Paper II were  
$2 \leq k  \leq  256$ and $L/k = 128, 256, 384, 512, 640$. The results are shown in Fig. \ref{fig2}: solid red (dark gray in grayscale) symbols and solid black symbols represent standard and inverse percolation thresholds for straight rigid $k$-mers on triangular lattices, respectively. The figure also shows the jamming curves corresponding to standard ($\theta_{j,k}$ vs $k$, open circles) and inverse ($\theta^i_{j,k}$ vs $k$, open squares) problems \cite{JSTAT6,JSTAT8}. 

Figure \ref{fig2} includes values of the percolation thresholds that were not reported in Papers I and II. Since one of the main objectives of this paper is to explore the relationship between the standard and inverse percolation processes in a triangular geometry, we calculated the corresponding percolation thresholds for a broader range of $k$. The new values, that reach $k = 512$, also give the possibility to better establish the maximum and minimum of the $\theta_{c,k}$[$\theta^i_{c,k}$] dependence with $k$ and to calculate the limit for $k=\infty$. 

For clarity, Fig. \ref{fig2} is divided into two data groups: (a) $1 \leq k \leq 40$, and (b) $16 \leq k \leq 512$. Solid diamonds and solid stars correspond to results from Papers I and II, respectively. Solid circles and solid squares indicate values obtained in the present work. In the case of $k=10,11,13,14,20,24$, the values of $\theta_{c,k}$ (standard percolation) were obtained by following the procedure and lattice sizes used in Paper I: $L/k=32,40,50,75,100$. In the case of $k=256, 340, 512$, two relatively small values of $L/k$ were used to calculate $\theta_{c,k}$ and $\theta^i_{c,k}$ ($L/k = 40$ and $L/k = 50$), with an effort reaching almost the limits of our computational capabilities. In all cases, $r=10^5$ computational runs were performed for each concentration $\theta$, on each lattice size $L$, and for each $k$-mer size $k$.

\begin{figure}
\includegraphics[width=0.9\columnwidth]{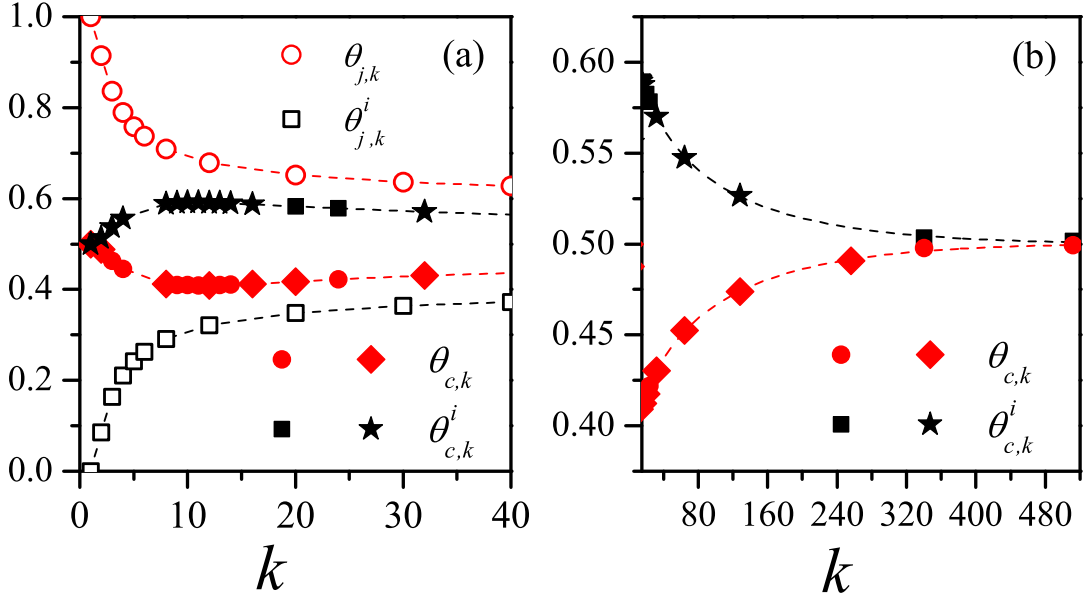}
\caption{(a) Standard (solid circles and solid diamonds) and inverse (solid squares and solid stars) percolation thresholds for straight rigid $k$-mers with $k$ ranging between 1 and 40 on triangular lattices. The figure also includes the jamming data corresponding to standard (open circles) and inverse (open squares) problems. Red (dark gray in grayscale) symbols and black symbols represent standard and inverse data. The curves were obtained by following the isotropic deposition/removal scheme. Solid diamonds and solid stars denote results from Papers I and II, respectively. Solid circles and solid squares indicate values obtained in the present work. (b) Standard and inverse percolation thresholds for values of $k$ varying between 16 and 512. Symbols are as in part (a). The dashed lines corresponds to the fitting curve for each system, from which it is get that $\theta_{c,k \rightarrow \infty} = 0.500(2)$ and $\theta^i_{c,k \rightarrow \infty} = 0.500(1)$. }  \label{fig2}
\end{figure}

The new results complement those previously reported in Papers I and II, and allow a deeper characterization of the  percolation transition occurring in triangular lattices. Thus, several important conclusions can be drawn from the data in Fig. \ref{fig2}.

Firstly, a complementarity property between the percolation thresholds for standard and inverse percolation is found: $\theta_{c,k} + \theta^i_{c,k} = 1$. This property is exact for the case $k=1$ \cite{Stauffer,Sykes} and, as shown in Fig. \ref{suma}, it holds for all the range of $k$, even in the limit of large values of $k$. In the figure, both simulation percolation thresholds were summed for each $k$. In every case,  $\theta_{c,k} + \theta^i_{c,k} = 1$ within the numerical error.  

\begin{figure}
\includegraphics[width=0.9\columnwidth]{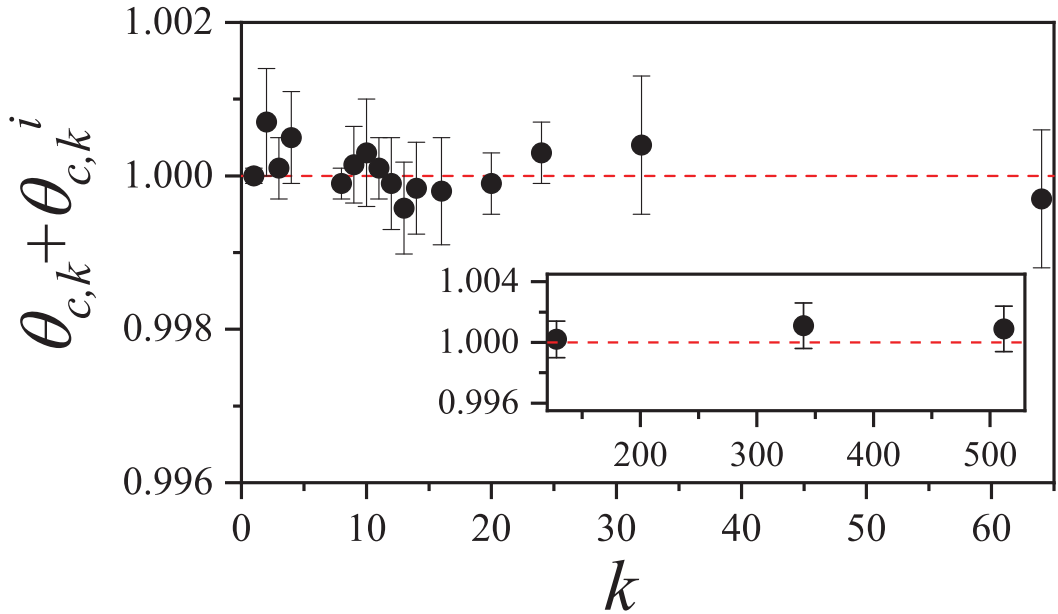}
\caption{The figure shows the sum of the standard and inverse percolation thresholds for all the range of considered $k$ sizes. The values correspond to the isotropic deposition/removal problem. As it can be observed, $\theta_{c,k} + \theta^i_{c,k} = 1$. }  \label{suma}
\end{figure}

The complementarity property is a nontrivial property and seems to be strongly dependent on the topology of the lattice. In fact, it is not observed for other systems, such as square \cite{Slutskii,JSTAT2} or honeycomb lattices \cite{JSTAT11,PRE27}. What is more, this property was not found in square bond lattices neither \cite{PRE16}, even when triangular site lattices and square bond lattices share the same coordination number. 

Secondly, the standard percolation threshold exhibits a non-monotonous dependence on the size $k$. $\theta_{c,k}$ decreases for small particles sizes, goes through a minimum around $k = 11$ [being $\theta_{c,k = 11} = 0.4091(3)$], and finally increases and asymptotically converges towards a definite value for large segments. The precise determination of this  minimum is reported here for the first time. In the case of inverse percolation, a maximum at $k = 11$ had been found in Paper II. This finding confirms the complementarity property discussed above.

Finally, the behavior of the percolation thresholds for large values of $k$ [see Fig. \ref{fig2}(b)] indicates that both 
$\theta_{c,k}$ and $\theta^i_{c,k}$ tend to 0.5 for infinitely long $k$-mers. In fact, the simulation data can be very well fitted with the functions $a_1\exp(-k/b_1)+a_2\exp(-k/b_2)+a_3$ and $a^i_1\exp(-k/b^i_1)+a^i_2\exp(-k/b^i_2)+a^i_3$ for standard and inverse percolation, respectively. In this case, the adjustment was performed for $k \geq 14$. The obtained results show that $a_3=\theta_{c,k \rightarrow \infty}=0.500(2)$ and $a^i_3=\theta^i_{c,k \rightarrow \infty}=0.500(1)$. In addition, $a_1 = -0.040(6)$, $a_2=-0.073(7)$, $b_1=30(5)$, and $b_2=120(10)$  for the standard case and  $a^i_1 =0.043(9)$, $a^i_2=0.070(9)$, $b^i_1=34(9)$, and $b^i_2=127(10)$ for the inverse percolation problem. As expected, the fitting curves fulfill the complementary condition $\theta_{c,k}+\theta^i_{c,k}=1$.

The limit value obtained here $\theta^i_{c,k \rightarrow \infty}=0.500(1)$ improves previous estimate in Paper II, where the value obtained of $\theta^i_{c,k \rightarrow \infty}$ was 0.506(2). Due to the lattice sizes used in this contribution, our present determination of $\theta^i_{c,k \rightarrow \infty}$ is expected to be more accurate than that reported previously.

In the case of standard percolation, the present results reveal a similar behavior to that reported for square lattices, where the percolation threshold tends asymptotically to a definite value for infinitely long $k$-mers \cite{Slutskii}. This  contrasts with the predictions in Paper I, which indicated (1) an increasing trend for $\theta_{c,k}$ at large values of $k$, and (2) the existence of a limit value $k \simeq 10^4$ from which all jammed configurations are nonpercolating states and, consequently, percolation would no longer occur. 

The new findings in this study (especially the complementarity property discussed above) provide a more complete and precise characterization of the standard percolation problem of straight rigid rods on triangular lattices. Namely, the $\theta_{c,k}$ curve divides the space of allowed values of $\theta$ in a percolating region and a non-percolating region. These phases extend to infinity in the space of the parameter $k$ so that the model presents percolation transition in all the range of $k$.  The existence of the percolation transition for the whole range of $k$ is consistent with the behavior observed in square lattices, in which jammed configurations reached by the deposition of needles always percolate \cite{kondrat2017}.

\subsection{Exact counting of configurations on finite cells}\label{cluster}

\begin{table}
\label{T1}
\begin{center}
\caption{Values of the quantities $T_l$, $C^{D}_l$ and $C^{I}_l$ for $4 \times 4$ lattices.}
\begin{tabular}{|c | c | c | c |  }
\hline
n & $T_l$ & $C^{D}_l$ & $C^{I}_l$ \\
\hline
\hline
0 & 1 & 0 & 1   \\ \hline
2 & 33 & 0 & 33   \\ \hline
4 & 412 & 20 & 392   \\ \hline
6 & 2485 & 585 & 1900   \\ \hline
8 & 7664 & 4416 & 3248   \\ \hline
10 & 11747 & 10321 & 1426   \\ \hline
12 & 7973 & 7901 & 72   \\ \hline
14 & 1802 & 1802 & 0   \\ \hline
16 & 56 & 56 & 0   \\ \hline
\end{tabular}
\end{center}
\end{table}

\begin{table}
\label{T2}
\begin{center}
\caption{Values of the quantities $T_l$, $C^{D}_l$ and $C^{I}_l$ for $5 \times 5$ lattices.}
\begin{tabular}{|c | c | c | c |  }
\hline
n & $T_l$ & $C^{D}_l$ & $C^{I}_l$ \\
\hline
\hline
0 & 1 & 0 & 1   \\ \hline
2 & 56 & 0 & 56   \\ \hline
4 & 1325 & 0 & 1325   \\ \hline
6 & 17384 & 386 & 16998   \\ \hline
8 & 139581 & 14180 & 125401   \\ \hline
10 & 714510 & 192618 & 521892   \\ \hline
12 & 2357344 & 1211811 & 1145533   \\ \hline
14 & 4957616 & 3755572 & 1202044   \\ \hline
16 & 6429895 & 5898574 & 531321   \\ \hline
18 & 4834116 & 4759098 & 75018   \\ \hline
20 & 1889380 & 1887961 & 1419   \\ \hline
22 & 313128 & 313128 & 0   \\ \hline
24 & 13872 & 13872 & 0   \\ \hline
\end{tabular}
\end{center}
\end{table}

\begin{table}
\label{T3}
\begin{center}
\caption{Values of the quantities $T_l$, $C^{D}_l$ and $C^{I}_l$ for $6 \times 6$ lattices.}
\begin{tabular}{|c | c | c | c |  }
\hline
n & $T_l$ & $C^{D}_l$ & $C^{I}_l$ \\
\hline
\hline
0 & 1 & 0 & 1   \\ \hline
2 & 85 & 0 & 85   \\ \hline
4 & 3226 & 0 & 3226   \\ \hline
6 & 72367 & 112 & 72255   \\ \hline
8 & 1070675 & 11697 & 1058978   \\ \hline
10 & 11040975 & 445881 & 10595094   \\ \hline
12 & 81784784 & 8733484 & 73051300   \\ \hline
14 & 442056227 & 99382990 & 342673237   \\ \hline
16 & 1753845586 & 691161330 & 1062684256   \\ \hline
18 & 5097923676 & 3000151582 & 2097772094   \\ \hline
20 & 10757573387 & 8227928526 & 2529644861   \\ \hline
22 & 16203594367 & 14427516941 & 1776077426   \\ \hline
24 & 16968630295 & 16294784319 & 673845976   \\ \hline
26 & 11881028004 & 11760742642 & 120285362   \\ \hline
28 & 5248329234 & 5240762986 & 7566248   \\ \hline
30 & 1337245213 & 1337175475 & 69738   \\ \hline
32 & 169111110 & 169111110 & 0   \\ \hline
34 & 7902376 & 7902376 & 0   \\ \hline
36 & 56568 & 56568 & 0   \\ \hline
\end{tabular}
\end{center}
\end{table}

An exact counting of configurations on finite cells was performed in order to back up the simulation predictions. This type of approach has been successfully applied to a variety of percolation problems \cite{PHYSA32,PHYSA34,PRE11}. Specifically, we will explore the relationship between standard and inverse thresholds from an analytical approach. The system chosen for the study was a RSA of dimers (objects occupying two consecutive lattice sites) on triangular lattices. The dimer is the simplest case of a straight rigid $k$-mer and contains all the properties of the multisite-occupancy deposition. 

We assume that the deposition or removal of dimers takes place on a small lattice of $m=l \times l$ sites. Once deposited the dimer remains ``frozen" on the substrate without dissociations or migrations. As $n$ dimers are deposited, the coverage is $\theta = 2n/m$. On the other hand, in the inverse case, as $n$ dimers are removed from an initially fully occupied lattice, the coverage is $\theta = 1-2n/m$. Thus, for any given $\theta$, different combinations of the $n$ dimers are possible each one of which will be called a configuration.

It is useful now to define the probability $r^{D[I]}_l(n)$ that a lattice composed of $l \times l$ sites percolates at a given value of $n$. The index $D[I]$ in the definition of $r_l$ indicates that $n$ is the number of deposited[removed] dimers. Then, for each value of $n$, $r^{D[I]}_l$ can be obtained as the ratio between the configurations that present a percolation cluster $C^{D[I]}_l$, and the total number of ways of distributing (depositing or removing) $n$ dimers on the $l \times l$ lattice $T_l$: $r^{D[I]}_l=C^{D[I]}_l/T_l$.

The quantities $T_l$, $C^{D}_l$ and $C^{I}_l$ for different values of $n$ are compiled in Tables I (case $l=4$), II (case $l=5$) and III (case $l=6$).  An own computer algorithm was developed to exactly calculate $T_l$, $C^{D}_l$ and $C^{I}_l$.

By observing Tables I-III, it is clear that $T_l(n)=C^{D}_l(n)+C^{I}_l(n)$ and, consequently, $r^D_l(n)+r^I_l(n)=1$. This finding is further proof that $\theta_c+\theta^i_c=1$, as found in previous section. This property is observed only for triangular lattices. In the case of square and honeycomb lattices, $T_l(n) \neq C^{D}_l(n)+C^{I}_l(n)$ and $\theta_c+\theta^i_c \neq 1$ (data not shown here for brevity).

\section{Percolation of straight rigid rods nematically deposited on triangular lattices}\label{nematic}

To have a more complete insight of the percolation processes in the triangular lattice, in this Section, the nematic percolation is studied with a focus in the complementary property observed in the isotropically case. 
For standard percolation, we reproduced and extended the results in Paper III. The results for inverse nematic percolation are reported for the first time.

\subsection{Model and basic definitions}\label{modnema}

To study the effect of $k$-mer alignment on percolation, straight rigid rods are deposited randomly, sequentially, and irreversibly on a $M = L \times L$ sites rhombus-shaped triangular lattice. The deposition process is performed as in Section \ref{modiso} but, now, the following restriction is considered: the $k$-mers are deposited along only one of the directions of the lattice. This leads to the formation of an oriented structure as depicted in Fig.\ref{fig4}(a). Periodic boundary conditions are considered in the deposition procedure. 

In order to distinguish between isotropic and nematic problem, for the rest of the paper we will use the variable $\vartheta$ to denote the concentration of occupied sites for the case of nematic deposition.

\begin{figure}
\includegraphics[width=0.95\columnwidth]{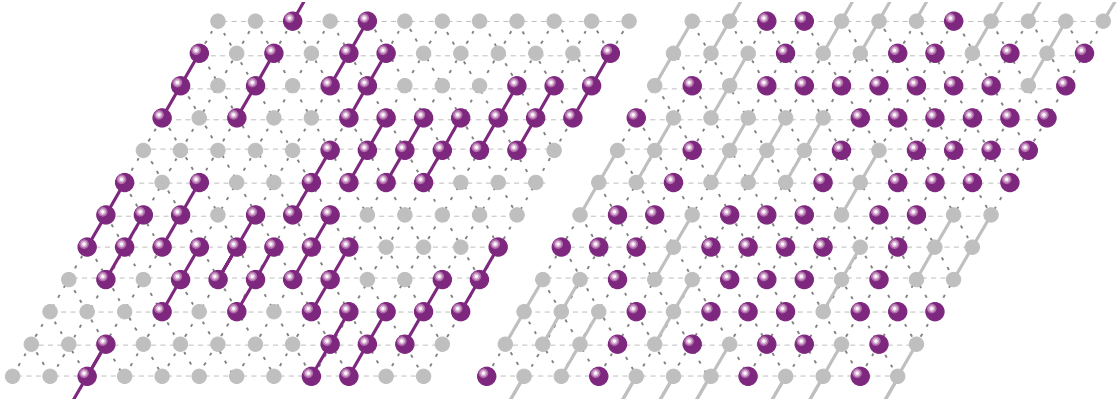}
\caption{(a) Schematic representation of a typical configuration obtained by nematic deposition of  $3$-mers ($k=3$) on a $L \times L$ lattice with $L=12$. Solid spheres joined by lines represent the deposited $k$-mers and gray circles correspond to empty sites. (b) Schematic representation of a typical configuration obtained by removing aligned $3$-mers from an initially fully occupied $L \times L$ lattice with $L=12$. Solid spheres represent occupied sites and gray circles indicate the empty sites resulting from the removal of the $k$-mers. Periodic boundary conditions are considered in parts (a) and (b).}  \label{fig4}
\end{figure}

The inverse percolation problem is also considered for the nematic case. We start from a fully occupied lattice that is diluted as follows: (1) one lattice direction $x_i \equiv \{x_1,x_2,x_3\}$ is chosen for the whole removal process; (2) a set of $k$ consecutive nearest-neighbor sites (aligned along the selected $x_i$ direction) is randomly chosen; and (3) if the $k$ sites selected in step (2) are occupied sites, then a $k$-mer is removed from those sites. Otherwise, the attempt is rejected. Steps (2) and (3) are repeated until the desired number of $k$-mers $N$ is removed from the lattice and the concentration of occupied particles is  $\vartheta^i=(M - kN)/M$.  The removal process leads to configurations as depicted in Fig.~\ref{fig4}(b). Periodic boundary conditions are considered.

The standard and inverse percolation thresholds are obtained thorough the extrapolation given by Eqs. (\ref{extrapolationdir}) and (\ref{extrapolationinv}). In this case, the equations can be written as
\begin{equation}
\vartheta_{c,k}(L)= \vartheta_{c,k} + \tilde{A}_k L^{-1/\nu},
\label{extrapolationdirnem}
\end{equation}
and
\begin{equation}
\vartheta^i_{c,k}(L)= \vartheta^i_{c,k} + \tilde{A}^i_k L^{-1/\nu},
\label{extrapolationinvnem}
\end{equation}
where $\tilde{A}_k$ and $\tilde{A}^i_k$ are the scaling constants for the standard and inverse nematic problem, respectively. Once the positions $\vartheta_{c,k}(L)$ and $\vartheta^i_{c,k}(L)$ are determined from the percolation probability functions $R_{L,k}(\vartheta)$, the percolation thresholds $\vartheta_{c,k}$ and $\vartheta^i_{c,k}$ can be obtained using the extrapolation scheme in Eqs. (\ref{extrapolationdirnem})-(\ref{extrapolationinvnem}). 

The obtained curves for $\vartheta_{c,k}$ and $\vartheta^i_{c,k}$ as functions of size $k$ are shown in Fig. \ref{fig8}. For both cases, a broad range of $k$ was studied ($1 \leq k \leq 512$): $1 \leq k \leq 128$, part (a); and $64 \leq k \leq 512$, part (b). Solid red (dark gray in grayscale) symbols and solid black symbols represent standard and inverse percolation thresholds, respectively. As in Fig. \ref{fig2}(a), Fig. \ref{fig8}(a) includes the jamming curves corresponding to standard ($\vartheta_{j,k}$ vs $k$, open circles) and inverse ($\vartheta^i_{j,k}$ vs $k$, open squares) problems. 

For $k$ between $2$ and $32$ lattices sizes $L/k = 128, 256, 384, 512$ and $640$ were considered; for $k$ between $64$ and $90$ $L/k = 32, 64, 128, 200$ and $256$; and for higher values of $k$, $k=128, 256, 300, 400, 512$, $L/k= 50$ up to $L/k = 150$. For $\vartheta_{c,k}$, the values obtained in the present work (solid circles) are consistent with those reported in Paper III \footnote{Previously published results for $k=64$ and $k=80$ are not shown in Fig. \ref{fig8}, and were replaced by new values calculated using a better statistics.}. (solid diamonds). On the other hand, in the case of the inverse percolation problem, the behavior of $\vartheta^i_{c,k}$ in terms of $k$  is reported here for the first time (solid squares in Fig. \ref{fig8}). 

\begin{figure}
\includegraphics[width=0.9\columnwidth]{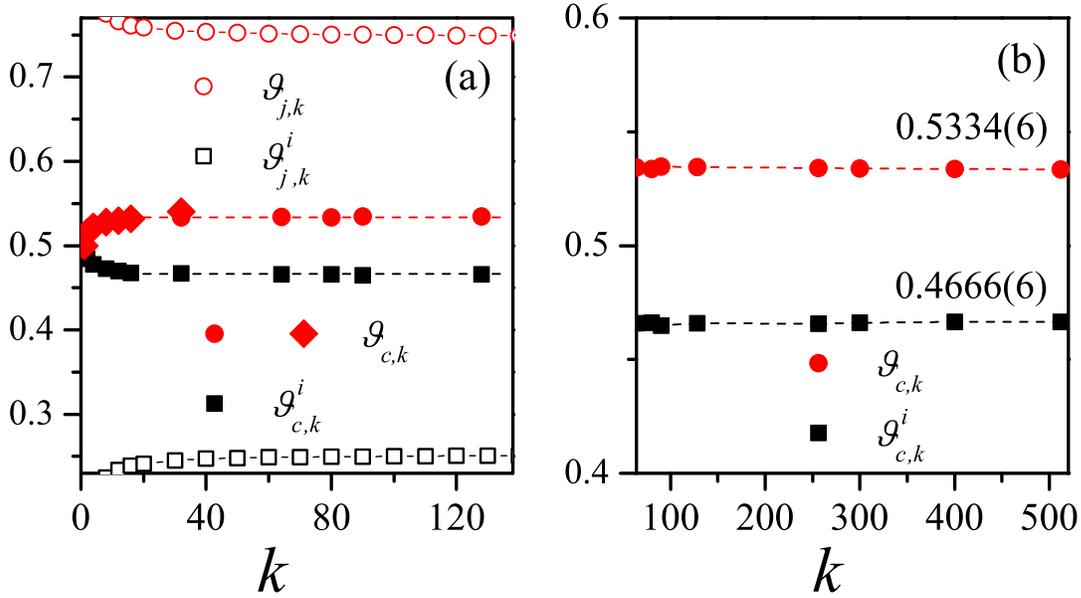}
\caption{(a) Standard (solid circles and solid diamonds) and inverse (solid squares) percolation thresholds for straight rigid $k$-mers with $k$ ranging between 1 and 128 on triangular lattices. The figure also includes the jamming data corresponding to standard (open circles) and inverse (open squares) problems. Red (dark gray in grayscale) symbols and black symbols represent standard and inverse data. The curves were obtained by following the nematic deposition/removal scheme. Solid diamonds denote results from Paper III, respectively. Solid circles and solid squares indicate values obtained in the present work. (b) Standard and inverse percolation thresholds for values of $k$ varying between 64 and 512. Symbols are as in part (a). The dashed lines corresponds to the fitting curve for each system, from which it is get that $\theta_{c,k \rightarrow \infty} = 0.5334(6)$ and $\theta^i_{c,k \rightarrow \infty} = 0.4666(6)$. }  \label{fig8}
\end{figure}


The curves of standard and inverse percolation thresholds are symmetric to each other with respect to the line $\theta = 0.5$.  As in the isotropic percolation problem, $\vartheta_{c,k} + \vartheta^i_{c,k} = 1$ within the numerical error and, accordingly, the complementarity property is also valid for the nematic percolation problem. The sum $\theta_{c,k} + \theta^i_{c,k}$ is shown in Fig. \ref{sumanem} for the whole range of $k$ values studied here.

\begin{figure}
\includegraphics[width=0.9\columnwidth]{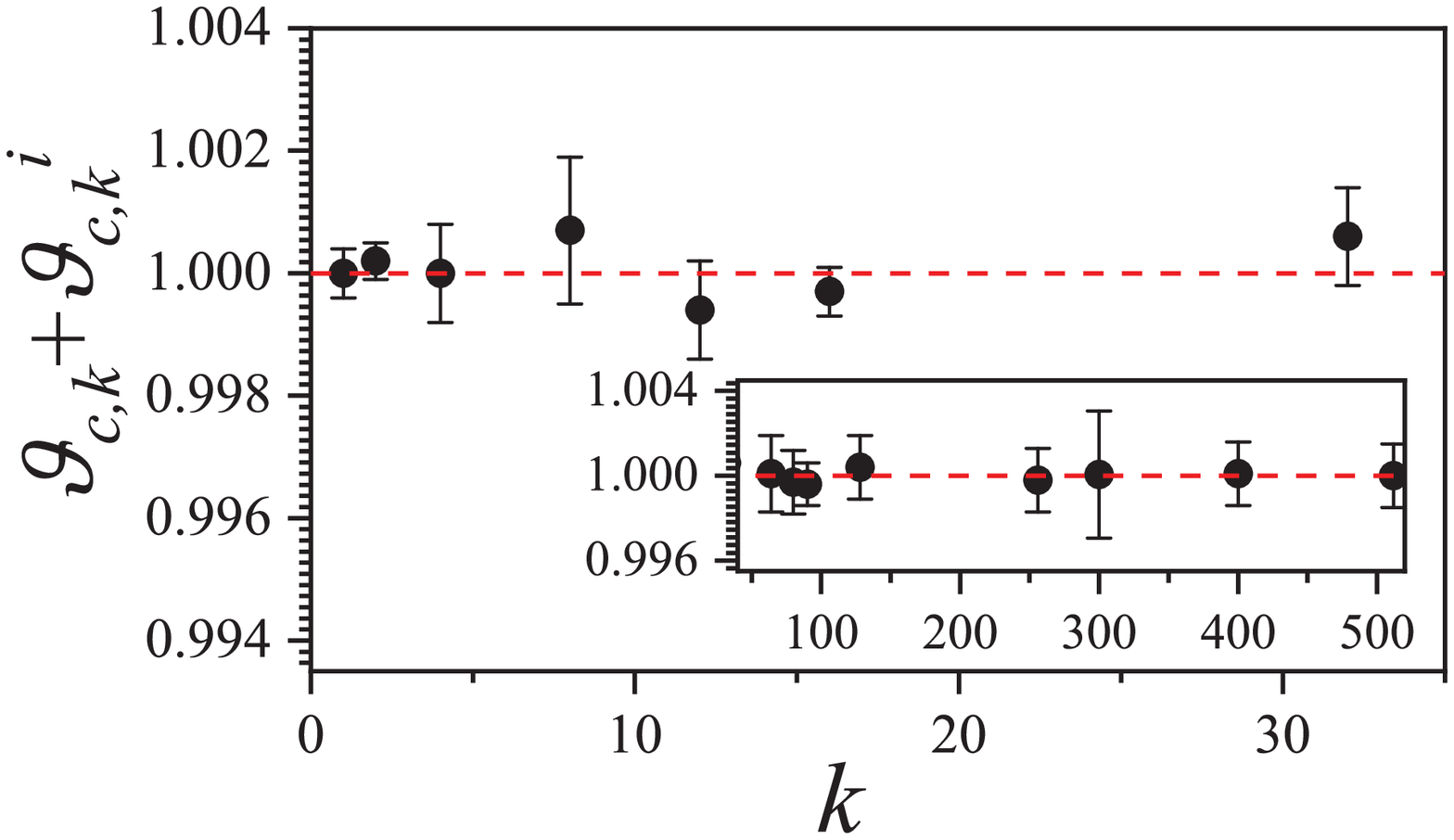}
\caption{ The figure shows the sum of the standard and inverse percolation thresholds for all the range of considered $k$ sizes. The values correspond to the nematic deposition problem. As it can be observed, $\theta_{c,k} + \theta^i_{c,k} = 1$. }  \label{sumanem}
\end{figure}


Standard and inverse percolation threshold show a monotonous dependence on the size $k$. $\vartheta_{c,k}$[$\vartheta^i_{c,k}$] rapidly increases[decreases] for small particles sizes and asymptotically converges towards a definite value for large segments. The behavior of the inverse percolation threshold $\vartheta^i_{c,k}$ as a function of $k$ is reported here for the first time.


The numerical data can be very well fitted with the functions $\tilde{a}_1\exp(-k/\tilde{b}_1)+\tilde{a}_2$ and 
$\tilde{a}^i_1\exp(-k/\tilde{b}^i_1)+\tilde{a}^i_2$ for standard and inverse percolation, respectively. The obtained results show that $\tilde{a}_2=\vartheta_{c,k \rightarrow \infty}=0.5334(6)$ and $\tilde{a}^i_2=\vartheta^i_{c,k \rightarrow \infty}=0.4666(6)$. In addition, $\tilde{a}_1 = -0.049(4)$,  and $\tilde{b}_1=2.8(4)$  for the standard case and  $\tilde{a}^i_1 =0.043(3)$, and $\tilde{b}^i_1=3.0(5)$ for the inverse percolation problem. These findings indicate that the RSA model of aligned $k$-mers on triangular lattices presents standard and inverse percolation transition in the whole range of $k$. As expected, the fitting curves satisfy the complementary condition $\vartheta_{c,k}+\vartheta^i_{c,k}=1$. 

The limit value obtained here $\vartheta_{c,k \rightarrow \infty}=0.5334(6)$ improves previous estimate in Paper III [$\vartheta_{c,k \rightarrow \infty}=0.582(9)$], where the standard nematic percolation threshold was calculated in the range $2 \leq k \leq 80$. The extension of the calculations to larger particles (in this case, up to $k=512$) led to a new and more accurate determination of $\vartheta_{c,k \rightarrow \infty}$. 


\begin{figure}
\includegraphics[width=0.8\columnwidth]{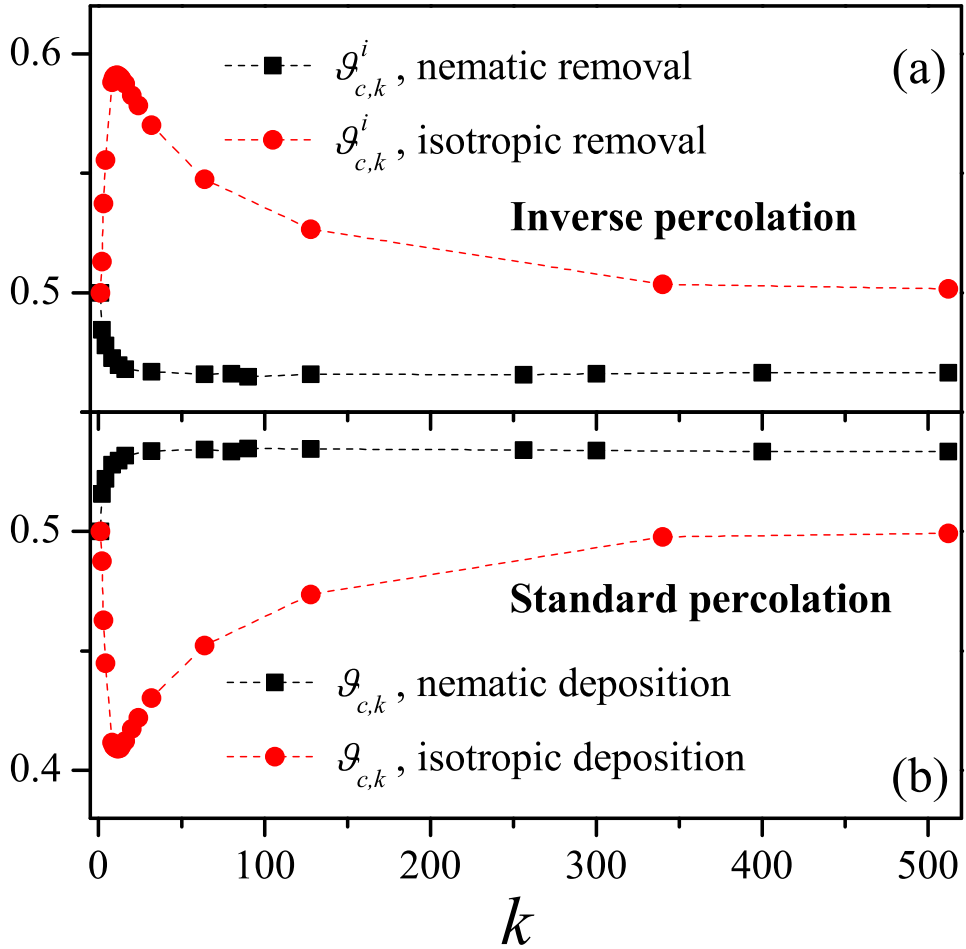}
\caption{(a) Comparison between isotropic (circles) and nematic (squares) percolation thresholds for the inverse percolation problem of straight rigid $k$-mers on triangular lattices. (b) Same as part (a) but for the standard percolation problem.}  \label{comp}
\end{figure}

To finish with the analysis of the percolation threshold curves, it is important to note that, for all $k$, the inverse percolation threshold of isotropic rods is higher than the corresponding one to aligned $k$-mers [see Fig. \ref{comp}(a)]. This is of interest since it means that is easier to disconnect the system when the needles are isotropically removed. In other words, the system is more robust when the removed needles are aligned in only one direction. This finding is consistent with the behavior observed for the standard percolation problem, in which the curve for nematic percolation is considerably above than the isotropic one in the whole range of $k$ [see Fig. \ref{comp}(b)]. Theoretical and experimental work support these predictions \cite{Du,Gross,Deng,Taherian}.

\section{Conclusions}
\label{conclusiones}

In this paper, standard and inverse percolation properties of straight rigid $k$-mers on triangular lattices have been studied by numerical simulations and finite-size scaling analysis. Two models have been addressed: isotropic model, where the deposition (removal) of the linear objects occurs with the same probability in any lattice direction; and nematic model, where one lattice direction is privileged for depositing (removing) the particles. 

For the isotropic deposition/removal problem, the previously reported results in Papers I and II were extended to longer $k$-mers (up to $k=512$). The standard and inverse percolation thresholds exhibit a non-monotonous dependence on the size $k$. $\theta_{c,k}$[$\theta^i_{c,k}$] decreases[increases] for small particles sizes, goes through a minimum[maximum] around $k = 11$, and finally increases and asymptotically converges towards a definite value for large segments. 

For large values of $k$ (after the minimum/maximum), the numerical data can be well fitted by the following functions: 
$\theta_{c,k}=a_1\exp(-k/b_1)+a_2\exp(-k/b_2)+a_3$ and $\theta^i_{c,k}=a^i_1\exp(-k/b^i_1)+a^i_2\exp(-k/b^i_2)+a^i_3$ ($k \geq 14$), being $a_1 = -0.040(6)$, $a_2=-0.073(7)$, $a_3=\theta_{c,k \rightarrow \infty}=0.500(2)$, $b_1=30(5)$, $b_2=120(10)$ and $a^i_3=\theta^i_{c,k \rightarrow \infty}=0.500(1)$.

The results obtained here allow us (1) to precisely determine the position of the minimum (maximum) observed in the curve of $\theta_c$ ($\theta^i_c$), located in $k=11$; (2) to establish the limit values $\theta_{c,k \rightarrow \infty}=0.500(2)$ and $\theta^i_{c,k \rightarrow \infty}=0.500(1)$. In the case of inverse problem, the present result corrects
the previously reported value of $\theta^i_{c,k \rightarrow \infty}=0.506(2)$ \cite{JSTAT8}; and (3) to conclude that $\theta_c+\theta^i_c=1$ for all value of $k$, even for infinitely long $k$-mers. This complementarity property was also validated by exact counting of configurations of dimers on finite cells.

The new findings provide a more complete and precise characterization of the percolation problem of straight rigid rods on triangular lattices. As occurs in the case of square lattices \cite{Slutskii,kondrat2017}, the $\theta_{c,k}$ curve divides the space of allowed values of $\theta$ in a percolating region and a non-percolating region. These phases extend to infinity in the space of the parameter $k$ so that the model presents percolation transition in all the range of $k$. This contrasts with the predictions in Paper I, which indicated the existence of a limit value $k \simeq 10^4$ from which all jammed configurations are nonpercolating states and, consequently, percolation transition is missed. 

Regarding the nematic case, the problem of aligned straight rigid $k$-mers deposited on triangular lattices was also revisited and extended. A increasing behavior was observed for $\vartheta_{c,k}$, with a finite value of saturation in the
limit of infinitely long $k$-mers: $\vartheta_{c,k}=\tilde{a}_1\exp(-k/\tilde{b}_1)+\tilde{a}_2$, being $\tilde{a}_1 = -0.049(4)$,  $\tilde{b}_1=2.8(4)$ and $\tilde{a}_2=\vartheta_{c,k \rightarrow \infty}=0.5334(6)$. This limit value improves previous estimate in Paper III [$\vartheta_{c,k \rightarrow \infty}=0.582(9)$], where the standard nematic percolation threshold was calculated in the range $2 \leq k \leq 80$.

In the case of inverse percolation by removing aligned $k$-mers from triangular lattices, the results of $\vartheta^i_{c,k}$ in terms of the size $k$ are presented for the first time in the literature: $\vartheta^i_{c,k}=\tilde{a}^i_1\exp(-k/\tilde{b}^i_1)+\tilde{a}^i_2$, with $\tilde{a}^i_1 =0.043(3)$, $\tilde{b}^i_1=3.0(5)$ and $\tilde{a}^i_2=\vartheta^i_{c,k \rightarrow \infty}=0.4666(6)$.

In both standard and inverse nematic problems, the obtained results indicate the existence of percolation phase transition in the whole range of $k$. In addition, and as in the isotropic case, the sum of standard and inverse percolation thresholds equals one ($\vartheta_{c,k}+\vartheta^i_{c,k}=1$) for all values of $k$, confirming the generality of this behavior in triangular lattices. Thus, the simple complementarity relationship between standard and inverse percolation thresholds seems to be a property typical for the triangular lattice, regardless isotropic/nematic deposition/removal. The complementarity property has not been observed in other regular lattices, showing that the lattice structure plays a fundamental role in determining the statistics and percolation properties of extended objects.

Finally, it was found that, for all $k$, the inverse percolation threshold of isotropic rods is higher than the corresponding one to aligned $k$-mers. It means that the phase of occupied sites is more robust when the removed sets of sites are aligned in only one lattice direction. A contrary behavior has been theoretically and experimentally observed for the standard percolation problem, where the curve for nematic percolation is above than the isotropic one in the whole range of $k$ \cite{Du,Gross,Deng,Taherian}.

\section*{Acknowledgements}

This work was supported in part by CONICET (Argentina) under project number PIP 112-201701-00673CO and Universidad Nacional de San Luis (Argentina) under project No. 03-0816. WL thanks support from Direcci\'on de Investigaci\'on Universidad de La Frontera (Chile), under projet DIUFRO No DI17-0024.

\newpage

\renewcommand{\baselinestretch}{1.0}


\end{large}

\end{document}